\documentclass[10pt,journal,compsoc]{IEEEtran}
\newif\ifpeerreview

\peerreviewtrue

\usepackage[nocompress]{cite}
\usepackage{url}
\usepackage{amsmath,amssymb,graphicx}
\usepackage{lipsum} 
\usepackage[switch]{lineno}

\usepackage{hyperref} 
\usepackage{xcolor,colortbl}
\usepackage{overpic}
\usepackage[nocompress]{cite}
\usepackage{amsmath,amssymb}
\usepackage{multirow}
\usepackage{multicol}
\usepackage{tabularx}
\usepackage{subcaption}
\usepackage{textcomp}
\usepackage{lipsum} 
\usepackage{color}
\usepackage{xspace}
\usepackage{booktabs} 
\usepackage[switch]{lineno}
\usepackage{transparent}

\def\eg{\emph{e.g.}\@\xspace}

\def\ie{\emph{i.e.}\@\xspace} 
\def\vs{\emph{vs.}\@\xspace}
\def\etc{\emph{etc.}\@\xspace}

\newcommand{\figLabel}{Fig.\xspace}

\newcommand{\secLabel}{Sec.\xspace}

\newcommand{\mysection}[1]{\vspace{3pt}\noindent\textbf{#1}}

\usepackage{listings}

\definecolor{Highlight}{HTML}{fc8d62}  

\newcommand{\pipeline}{DN$\to$SR$\to$DM\xspace}
\newcommand{\pipelinejoint}{DN+SR$\to$DM\xspace}
\newcommand{\RR}[1]{{#1}}

\title{Rethinking \RR{Learning-based} Demosaicing, Denoising, and Super-Resolution Pipeline}

\author{
Guocheng Qian$^{1\ast}$,
Yuanhao Wang$^{1\ast}$, 
Jinjin Gu$^{2}$, 
Chao Dong$^{3,4}$, \\
Wolfgang Heidrich$^{1}$, 
Bernard Ghanem$^{1}$, 
Jimmy S. Ren$^{5, 6}$\\
\IEEEcompsocitemizethanks{\IEEEcompsocthanksitem 
$^{1}$KAUST,
$^{2}$The University of Sydney, 
$^{3}$Shanghai AI Laboratory, 
$^{4}$Shenzhen Institutes of Advanced Technology, Chinese Academy of Sciences
$^{5}$SenseTime Research, 
$^{6}$Qing Yuan Research Institute, Shanghai Jiao Tong University, 
\protect\\
\IEEEcompsocthanksitem $^{\ast}$Equal contribution.}
}

\begin{document}

\IEEEtitleabstractindextext{%
\begin{abstract}
    \RR{Imaging is usually a mixture problem of incomplete color sampling, noise degradation, and limited resolution. This mixture problem is typically solved by a sequential solution that applies demosaicing (DM), denoising (DN), and super-resolution (SR) sequentially in a fixed and predefined pipeline (execution order of tasks), DM$\to$DN$\to$SR.
    The most recent work on image processing focuses on developing more sophisticated architectures to achieve higher image quality. Little attention has been paid to the design of the pipeline, and it is still not clear how significant the pipeline is to image quality.
    In this work, we comprehensively study the effects of pipelines on the mixture problem of learning-based DN, DM, and SR, in both sequential and joint solutions. On the one hand, in sequential solutions, we find that the pipeline has a non-trivial effect on the resulted image quality. Our suggested pipeline DN$\to$SR$\to$DM yields consistently better performance than other sequential pipelines in various experimental settings and benchmarks. On the other hand, in joint solutions, we propose an end-to-end Trinity Pixel Enhancement NETwork (TENet) that achieves the state-of-the-art performance for the mixture problem. We further present a novel and simple method that can integrate a certain pipeline into a given end-to-end network by providing intermediate supervision using a detachable head. Extensive experiments show that an end-to-end network with the proposed pipeline can attain only a consistent but insignificant improvement. Our work indicates that the investigation of pipelines is applicable in sequential solutions, but is not very necessary in end-to-end networks. \RR{Code, models, and our contributed PixelShift200 dataset are available at \url{https://github.com/guochengqian/TENet}}.
}
\end{abstract}

}

\maketitle
\thispagestyle{empty}

\begin{figure*}[!t]
\begin{center}
\centering
{\parbox{1.0\textwidth}{\begin{overpic}[width=\linewidth]{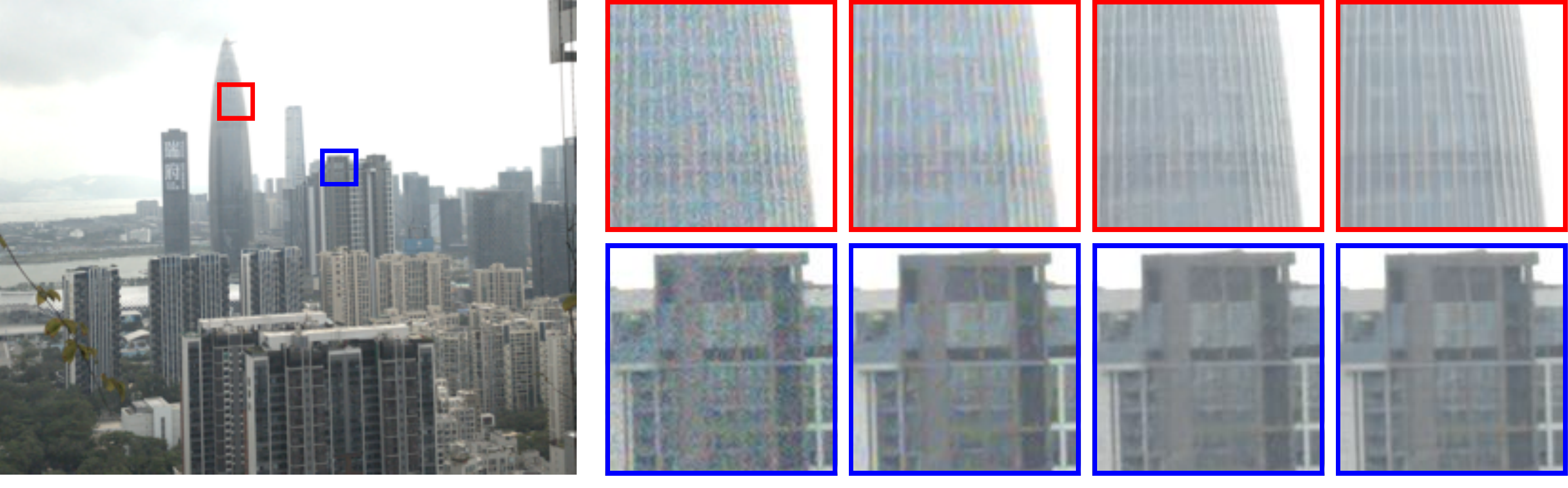}
\put(0.5,-1.5){\scalebox{1.0}{Real-World Image Captured by iPhone X}}
\put(43,-1.5){\scalebox{1.0}{DCRaw}}
\put(56.5,-1.5){\scalebox{1.0}{Camera Raw}}
\put(72.5,-1.5){\scalebox{1.0}{JDnDmSR\cite{XingEndtoEndLF}}}		\put(86.5,-1.5){\scalebox{1.0}{TENet++ (ours)}}

\end{overpic}}}
\vspace{1em}
\caption{\textbf{Qualitative comparisons for joint DM, DN, and SR ($\times 2$) on a real raw image captured by an iPhone X.}  Our TENet++ delivers a more visually appearing result compared to  popular software DCRaw and Camera Raw and state-of-the-art JDnDmSR \cite{XingEndtoEndLF}, producing less color distortions and more fine-grained details. The output of DCRaw and Camera Raw is superresolved by a SR model implemented by \RR{the same 6 RRDB blocks} as TENet++ for a fair comparison (\secLabel \ref{sec:exp}).} 
\label{fig:teaser}
\end{center}
\end{figure*}

\IEEEraisesectionheading{
\section{Introduction}\label{sec:introduction}}
\IEEEPARstart{O}{btaining} high-quality, high-resolution images has attracted increasing attention. Acquiring such images is difficult in practice due to hardware limitations, especially for mobile devices.
First, most digital cameras capture images using a single image sensor overlaid with a color filter array (\eg Bayer pattern), which causes incomplete color sampling, \ie resulting in mosaic images instead of RGB images.
Second, images taken directly from the image sensor are inevitably noisy.
Third, typical mobile devices are equipped with limited pixel numbers and lenses with fixed and short focal lengths, which makes imaging of distant or small objects challenging and limits image resolution. The real-shot image captured by an iPhone X shown in \figLabel \ref{fig:teaser} shows unnatural colorization, noise, and loss of detail due to these limitations.
Demosaicing (DM) \cite{Kimmel1999DemosaicingIR}, denoising (DN) \cite{dabov2007bm3d} and super-resolution (SR) \cite{Irani1991ImprovingRB} are the three fundamental tasks that have been studied and included in image processing pipelines (ISPs\footnote{ISP can be the abbreviation for image processing pipeline or image signal processor. We use these terms interchangeably.}) to resolve the hardware limitations mentioned above and to improve image quality. 

Deep learning technologies \cite{gharbi2016deep,zhang2017beyond,SRCNN} have recently led to breakthrough progress in DN, DM, and SR algorithms, \RR{and have spawned commercial products using learning-based image processing such as modern mobile phones (iPhone, Google Pixel, \etc)}.  
Despite the achievement of \RR{deep learning} in each task, \textit{imaging is usually a mixture problem of incomplete color sampling, noise degradation, and resolution limitation}. The combination of DN, DM, and SR is more common and more complicated than any single problem in practical application.

Previous methods handle the mixture problem through a sequential solution that performs DM, DN, and SR independently in a predefined and fixed order: DM$\to$DN$\to$SR \cite{Nakamura2005ImageSA}, \ie firstly DM, followed by DN, and then SR. \RR{Recent methods instead show a trend in performing DN and SR in mosaic space before DM \cite{Chatterjee2011NoiseSI,Brooks2019UnprocessingIF,Xu2019TowardsRS}}. 
However, these works do not consider the important but under-explored mixture problem of DN, DM, and SR. Furthermore, it is not clear how significant the execution order of tasks (\ie pipeline) is to the performance of this mixture problem. 

In this paper, we analyze the characteristics of DN, DM, and SR and the behaviors of their interactions. We find that issues caused by interactions between tasks occur when the corresponding algorithms are applied sequentially to solve the mixture problem. For example, superresolving a demosaiced image will magnify artifacts (\eg moir\'e) introduced by the DM algorithm (see \figLabel \ref{fig:result_pipeline}). 
\textit{We propose a novel image processing pipeline: DN$\to$SR$\to$DM, for sequential solutions}. We find that the proposed pipeline can alleviate problems caused by task interactions to a great extent.
Extensive experiments of \RR{learning-based DN, DM, and SR} show that our pipeline can \textit{consistently} improve image quality \RR{of sequential solutions}, regardless of architecture, dataset, and SR factor (see \secLabel \ref{sec:ablation}).

\RR{We further study the effect of pipelines in joint solutions (end-to-end networks) for the mixture problem}. We first propose a \textbf{T}rinity \textbf{E}nhancement \textbf{Net}work (\RR{TENet++}\footnote{\RR{We add ++ after TENet to avoid confusion with the outdated architecture used in previous arXiv version of this paper}}) to address the mixture problem.
\RR{We then present a simple yet effective way that enforces an end-to-end network to follow a certain pipeline by providing intermediate supervision. Through experiments on TENet++ and other architectures \cite{Mei_2021_CVPR, zhou2018deep, XingEndtoEndLF}, we notice marginal but consistent improvements after inserting the proposed pipeline. Our studies suggest that the investigation of pipelines in end-to-end networks can improve the performance but is not very necessary considering the insignificant improvement. }

\mysection{Contributions:}
\textbf{(1)}
We are the first to propose and analyze the mixture problem of \RR{learning-based denoising, demosaicing, and super-resolution}.
\textbf{(2)} We suggest a new pipeline: DN$\to$SR$\to$DM for solving the mixture problem of DN, DM, and SR. Extensive experiments show that the proposed pipeline can consistently improve performance \RR{for sequential solutions}.
\textbf{(3)} We propose an end-to-end network named Trinity Pixel Enhancement Network (TENet++) that achieves SOTA performance for joint DN, DM, and SR.
\RR{\textbf{(4)} We show how to make an end-to-end network follow a certain pipeline. We indicate an insignificant effect of pipelines on end-to-end networks.}
\textbf{(5)} We notice that there is a lack of full-color sampled datasets in the literature. We contribute a new real-world dataset, namely \textit{PixelShift200}, which consists of red, green, and blue channels without the need for color interpolation. We demonstrate the benefits of PixelShift200 in training and evaluating raw image processing tasks. 
\looseness=-1

\section{Related Work}
\label{sec:related_work}

\mysection{Demosaicing.}
Digital cameras take subsampled color measurements at alternating pixel locations. The resulting images of the subsampled measurements are named mosaic images. The mosaic images are then interpolated to create full-color images with per-pixel red, green, and blue information by a so-called demosaicing (DM) process. Early DM methods are model-based \cite{malvar2004high,hirakawa2005adaptive,zhang2005color}, which focus on the construction of filters (\eg edge-aware interpolation) and image priors (\eg chrominance continuity). Model-based methods are still commonly used in camera systems and software; \eg the image processing library DCRaw utilizes \cite{hirakawa2005adaptive}. Pioneering works also explored data-driven methods \cite{kapah2000demosaicking,wang2014multilayer} that learn a mapping from a raw image to an RGB image. Recently, deep learning has achieved overwhelming performance in DM. \cite{gharbi2016deep} presented DemosaicNet, a deep convolutional neural network (CNN)-based DM algorithm that outperforms the previous methods by a large margin. 
Following DemosaicNet, many works  \cite{DBLP:conf/iclr/ZhangLLZF19,Bammey2020AnAN} design different architectures to improve the demosaicing quality.  

\mysection{Denoising.}
Noise is inevitable during the imaging process. Early denoising (DN) methods exploited image priors, such as content variance \cite{rudin1992nonlinear}, self-similarity
\cite{buades2005nonlocal}, and sparse representation \cite{aharon2006k} for image denoising. 
The most recent denoisers are entirely data-driven, consisting of CNNs trained to recover noisy images to noise-free targets \cite{burger2012image,zhang2017beyond}.
Despite the effectiveness of these learning-based denoisers on synthesized benchmarks \cite{BSD100}, they generalize poorly to real-shot images due to their oversimplified assumption that noise is additive, white, and Gaussian \cite{Pltz2018NeuralNN}. While the noise pattern of a color image is complex because of nonlinear image processing (DM, color mapping, and compression), the noise patterns on raw images are well studied. \cite{Hasinoff2014PhotonPN}
characterized how sensor noise primarily comes from two sources: Poisson noise (shot noise) and Gaussian noise (read noise). To improve the generalization ability of deep denoisers, \cite{Brooks2019UnprocessingIF,Zamir2020CycleISPRI} proposed denoising on raw images using Poisson-Gaussian noise, which outperformed previous methods on the real-world image denoising dataset DND \cite{Plotz2017BenchmarkingDA}.
In this paper, we find that denoising RAW images directly yields higher quality, regardless of the network architecture. 

\mysection{Super-resolution.}
Due to the limited sensor size, image resolution is usually not as high as desired.
Image SR aims to recover a high-resolution (HR) image from its low-resolution (LR) version.
Previously, example-based SR methods \cite{Glasner2009SuperresolutionFA, A+} that exploit the self-similarity property provided state-of-the-art performance. 
Recently, learning-based methods \cite{SRCNN, FSRCNN} developed the CNN-based SR algorithms SRCNN and FSRCNN, outperforming example-based methods. After these seminal works, many learning-based SR methods have emerged \cite{ESRGAN, Mei_2021_CVPR}. However, most of them focus on color image SR. Only a few works have paid attention to the SR of raw images \cite{zhou2018deep, Xu2019TowardsRS, Liu2021ExploitCR}. 

\mysection{ISP and Mixture Problem.}
Image processing is always accompanied by a mixture problem of DN, DM and SR. An ISP is embedded in a modern camera to perform all these tasks. Most ISPs solve tasks independently and sequentially through the predefined pipeline DM$\to$DN$\to$SR \cite{Nakamura2005ImageSA}.
\RR{Although some previous works proposed new pipelines, such as performing DN before DM \cite{Chatterjee2011NoiseSI,Brooks2019UnprocessingIF}}, less attention has been paid to the execution order of joint DN, DM, and SR, especially since many ISP methods in the deep learning era are now learning-based and leverage end-to-end algorithms \cite{flexisp, zhou2018deep,  Schwartz2019DeepISPTL, Xu2019TowardsRS, XingEndtoEndLF}. These end-to-end solutions map a raw image to a desired RGB image directly, without focusing on the pipeline. In this work,  we diverge from the common architecture engineering in the area of image processing and rethink the mixture problem of DM, DN and SR from a holistic perspective, and more especially, the execution order (pipeline) of tasks.   
\section{Methodology}\label{sec:method}
\subsection{A New Pipeline for DN, DM and SR}\label{sec:new_pipeline}
We propose a new image processing pipeline, DN$\to$SR$\to$DM, that significantly improves the image quality \RR{of sequential solutions} for the joint problem of DM, DN and SR.
For a given noisy LR raw image $M_n^{LR}$, our pipeline obtains the final HR color image $I^{HR}$ from $M_n^{LR}$ using a composite function as follows:
\begin{equation}\label{eqn:pipeline}
	I^{HR}
	=\mathcal{C}_M(\mathcal{S}_M(\mathcal{D}_M(M_n^{LR}))),
\end{equation}
where $\mathcal{C}_M$ is the demosaicing function ($\mathcal{C}$ denotes ``\textbf{c}olorize''), $\mathcal{S}_M$ the \textbf{S}R function for mosaic images, and $\mathcal{D}_M$ the \textbf{d}enoising function for mosaic images. $M$ and subscript $_M$ stand for ``mosaic'', while subscript $_n$ indicates noisy.
We first perform DN on the noisy raw mosaic image to obtain its noise-free version, $M^{LR}=D_M(M_n^{LR})$.
We then adopt $S_M$ to superresolve the LR mosaic image and obtain a HR mosaic image, $M^{HR}=\mathcal{S}_M(M^{LR})$.
Finally, we use DM to interpolate $M^{HR}$ to a full-color HR image, $I^{HR}=\mathcal{C}_M(M^{HR})$. 

\noindent\textbf{Why perform DN at the first stage}? \label{sec:dn_first}
DN is usually performed after DM in a typical ISP. We propose DN first for three reasons:
(1) the noise model for raw images has been well studied (Gaussian-Poisson distribution). The quality of DN is higher for raw images than for color images.
(2) The existence of noise adds complexity to subsequent tasks. Noise has a high possibility of hiding color information and destroying textures, depending on the noise level. Processing a noisy image will result in unwanted artifacts in most cases. For example, \RR{\figurename~\ref{fig:result_pipeline} (DM$\to$DN$\to$SR) showcases that demosaicing a noisy image is prone to moir\'e.} 
(3) Image processing prior to DN will degrade the noise pattern and complicate denoising. For example, SR will destroy the noise distribution and make removal of noise from the super-resolved image extremely difficult. \RR{\figurename~\ref{fig:result_pipeline} (DM$\to$SR$\to$DN) shows such an example, where obvious noise appears.}

\noindent\textbf{Why perform SR before DM?} \label{sec:sr_before_dm}
Previous ISPs usually first demosaic a raw image into a color image and then perform SR. 
We suggest super-resolving the raw image to a higher resolution before conducting DM. \RR{In other words, super-resolution in our suggested pipeline is performed on mosaic images instead of RGB images.}
Our proposed pipeline has at least two advantages:
(1) demosaicing a higher resolution raw image yields fewer artifacts than demosaicing a lower resolution image. A DM algorithm usually introduces conspicuous artifacts (zippering, color moir\'e, and blurring) in the high-frequency texture regions, especially when the input resolution is low. These artifacts are alleviated when DM is applied to an image with higher resolution.
(2) The artifacts caused by super-resolving the defects of a demosaiced image can be avoided in our pipeline. 
\RR{As shown in \figLabel~\ref{fig:result_pipeline}, DN$\to$SR$\to$DM that performs SR before DM alleviates color distortion and moir\'e compared to its counterpart DN$\to$DM$\to$SR.}

\begin{figure*}[t]
\begin{center}
\begin{subfigure}[b]{0.45\textwidth}
\includegraphics[width=1.0\textwidth]{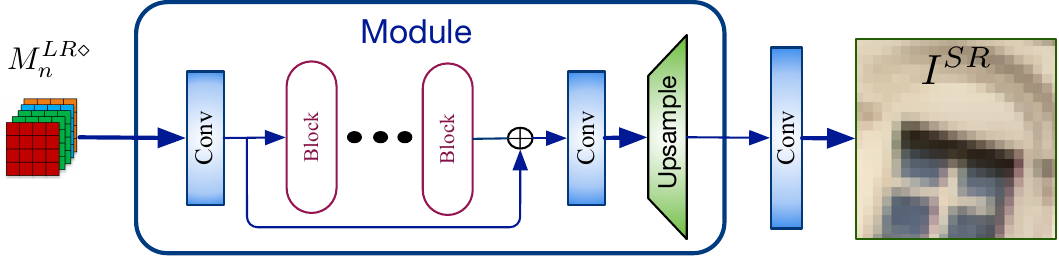}
\caption{E2ENet}
\label{fig:arch_e2e}
\end{subfigure}
\hfill
\begin{subfigure}[b]{0.45\textwidth}
\includegraphics[width=1.0\textwidth]{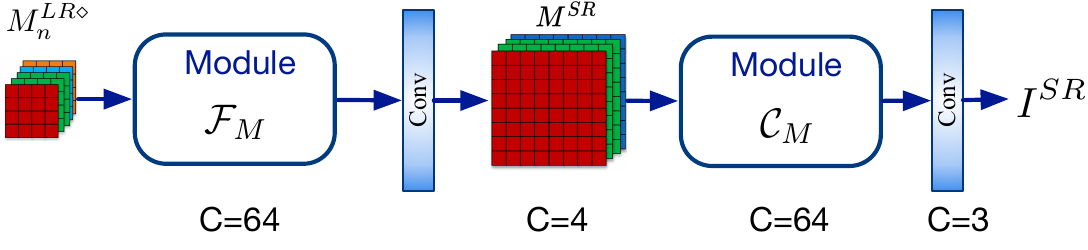}
\caption{TENet}
\label{fig:arch_tenet_old}
\end{subfigure}
\begin{subfigure}[b]{0.45\textwidth}
\includegraphics[width=1.0\textwidth]{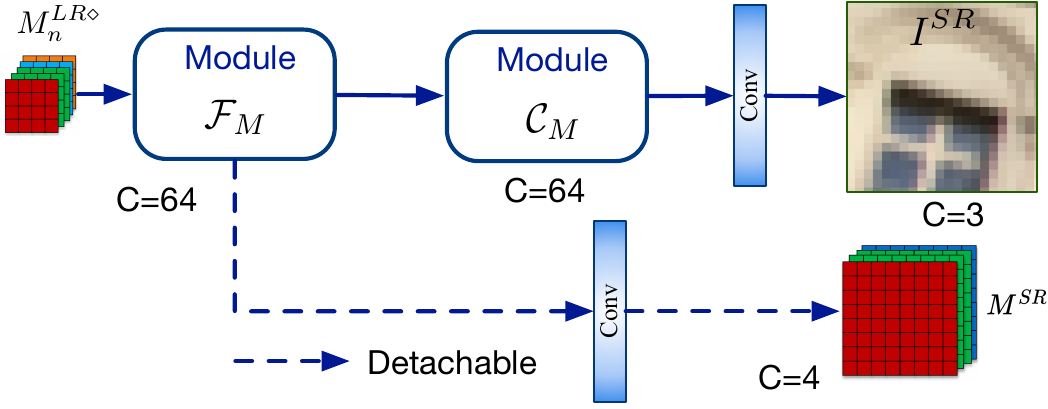}
\caption{TENet++}
\label{fig:arch_tenet}
\end{subfigure}
\hfill
\begin{subfigure}[b]{0.45\textwidth}
\includegraphics[width=0.9\textwidth]{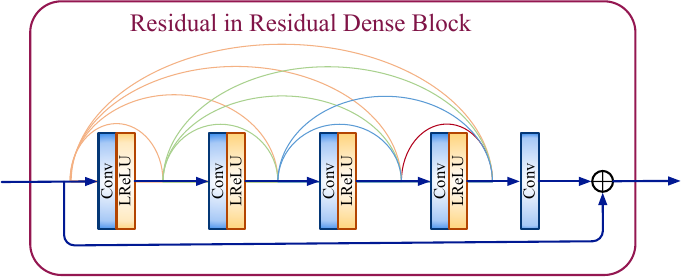}
\caption{residual in residual dense block (RRDB)}
\label{fig:arch_rrdb}
\end{subfigure}
\caption{
\RR{
\textbf{Architecture of our TENet++ (c). }
(a) E2ENet is a one-stage end-to-end network that learns a mapping from the noisy LR raw image to the HR color image directly using a single module.
(b) Our naive version of Trinity Enhancement Network, denoted as TENet,  consists of two main components: a joint denoising and super-resolution module $\mathcal{F}_M$ and a demosaicing module $\mathcal{C}_M$. Each module shares the same architecture as (a), which composes convolutional layers to extract features and an upsampling layer to interpolate features. This two-component design makes the network follow a certain pipeline (supersolve raw image before demosaicking) for the joint DN, DM, and SR problem, and facilitates optimization by providing intermediate supervision, compared to (a). 
However, TENet suffers from the bottleneck issue (channel size is dropped to $C=4$ in the middle of the network).
(c) Our proposed TENet++ where a detachable convolution layer is adopted after $\mathcal{F}_M$ for reconstructing the high-resolution raw image ($M^{SR}$). This detached layer is activated during training, thus eschewing TENet++ from the bottleneck issue, and is detached in inference. 
(d) The default block (RRDB \cite{ESRGAN}) used in TENet++. 
}
}
\label{fig:net}
\end{center}
\end{figure*}
\subsection{Inserting Our Pipeline into An End-to-end Network}\label{sec:pip_joint}

Despite the effectiveness of the proposed pipeline, simply performing multiple tasks sequentially and independently, as shown in Equation \ref{eqn:pipeline} reduces performance. 
For example, DN will introduce blurring in subsequent tasks.
An important reason for this performance drop is that no appropriate model can perfectly handle the intermediate state.
The intermediate state refers to the temporal result after previous processing and usually involves complex task-related defects that affect subsequent tasks. With the advent of deep learning-based methods, we can address complicated multitask problems in an end-to-end manner, \ie a ``joint solution''. 
Although the joint solution has shown impressive performance in a variety of tasks \cite{klatzer2016learning, zhang2018learning, Xu2020UnifiedDC}, it is still underexplored for joint DN, DM, and SR. 
The most recent works \cite{zhou2018deep, Xu2019TowardsRS, Liu2020JointDA, XingEndtoEndLF} focused on such a mixture problem.
However, most of them simply treat the whole network as a black box, without considering the pipeline inside. Their methods just learn a mapping from the noisy LR raw image to the HR color image, with the final target (the output of a camera ISP) serving as supervision.  We denote this type of one-stage end-to-end black-box network as \textbf{E2ENet}, \RR{whose architecture is illustrated in \figLabel \ref{fig:arch_e2e}.}
We denote E2ENet's pipeline as DN$+$SR$+$DM. 

We show how to make an end-to-end network follow a certain pipeline to solve the mixture problem instead of just learning a one-stage mapping. 
With the joint solution, we can \textit{simplify the sequential pipeline \pipeline as \pipelinejoint}.
Compared to E2ENet (DN$+$SR$+$DM), we assign a specific task to each component of the network. Our network performs joint DN and SR in the first stage, followed by DM in the final stage. We achieve this pipeline by providing intermediate supervision when training an end-to-end network.  
We denote the mapping function of joint DN and SR as $\mathcal{F}_M$, and the DM mapping as $\mathcal{C}_M$. 
$\mathcal{F}_M$ and $\mathcal{C}_M$ can be trained jointly.
The $l_1$-norm loss for the final output is calculated by:
\begin{equation}
	\mathcal{L}_{joint}=\|\mathcal{C}_M(\mathcal{F}_M(M_n^{LR}))-I^{HR}_{gt}\|,
\end{equation}
where $I^{HR}_{gt}$ represents the ground-truth HR color image of the input LR noisy raw image $M_n^{LR}$.
We further construct an intermediate output $M^{HR}$ (the superresolved mosaic image) and propose an intermediate loss, $\mathcal{L}_{SR}$, as follows: 
\begin{equation}
	\mathcal{L}_{SR}=\|\mathcal{F}_M(M_n^{LR})-M^{HR}_{gt}\|,
\end{equation}
where $M^{HR}_{gt}$ represents the ground-truth HR noise-free mosaic image and the output of $\mathcal{F}_M(M_n^{LR})$ is $M^{SR}$. 
The $\mathcal{L}_{SR}$ loss makes the first part of the network focus on joint DN and SR, and the second part on DM. The $\mathcal{L}_{joint}$ loss controls the fidelity of the final output. The final objective function is the sum of two loss terms: 
\begin{equation}
	\mathcal{L}=\mathcal{L}_{joint}+\mathcal{L}_{SR},
\end{equation}
\RR{While \pipeline outperforms other pipelines in sequential solutions, \pipelinejoint is the best overall in joint solutions (despite the marginal improvements). Note here adding additional denoising supervision is not beneficial to the performance as shown in our experiment. This demonstrates that the essence of our proposed pipeline is to perform DN and SR in mosaic space, not RGB space, which are our core arguments for both pipelines (see Sec. \ref{sec:new_pipeline}).}

\subsection{Trinity of Pixel Enhancement Network}\label{sec:network}
\RR{The naive solution to achieve the \pipelinejoint pipeline is to concatenate two subnetworks $\mathcal{F}_M$ and $\mathcal{C}_M$ in the network backbone and actually produce $M^{SR}$ in the middle of the network, as shown in \figLabel \ref{fig:arch_tenet_old}. This is the architecture that we used in the preprint version of our work and is denoted TENet. Unfortunately, this solution will face performance drops due to a bottleneck issue. The bottleneck arises as the channel size $C$ is decreased from the latent space (\eg $C=64$) to the raw image space ($C=4$) to yield the SR raw image.} To solve this issue, we present \textbf{T}rinity of Pixel \textbf{E}nhancement \textbf{N}etwork (\textbf{TENet++}). TENet++ leverages an attachable branch to provide additional supervision during training. 
The attachable branch is implemented by a single convolutional layer to map the feature from the latent space to the raw image space.
The architecture of TENet++ is illustrated in \figLabel \ref{fig:arch_tenet}. 
The noisy LR mosaic image $M_n^{LR}$ with size $H \times W$ is reshaped to a four-channel image (red, green, green, blue) with size $\frac{H}{2}\times \frac{W}{2} \times 4$.
The noise variance for each channel is concatenated into the reshaped raw image. The eight-channel input is denoted $M_n^{LR\diamond}$, which is passed to the TENet++ backbone. 
\RR{TENet++ consists of two components in its backbone: a joint denoising and super-resolution module $\mathcal{F}_M$ and a demosaicing module $\mathcal{C}_M$.
$\mathcal{F}_M$ and $\mathcal{C}_M$ share the same structure as the module used in E2ENet (detailed in \figLabel \ref{fig:arch_e2e}). Both $\mathcal{F}_M$ and $\mathcal{C}_M$ are composed of a convolution layer to transform features, $N/2$ convolutional blocks to extract features, and a convolution layer with an upsampling layer to interpolate features. Note $N$ is the total number of blocks in TENet++ and is set to $12$ by default. The upsampling ratio of $\mathcal{F}_M$ is the SR ratio ($2$ by default), while the upsampling ratio of $\mathcal{C}_M$ equals $2$ since $\mathcal{C}_M$ is the demosaicing module to interpolate colors.}
We employ the Residual in Residual Dense Block (RRDB) proposed in ESRGAN \cite{ESRGAN} (see \figLabel \ref{fig:arch_rrdb}) to implement the blocks used in each module by default. A pixel shuffle layer \cite{shi2016real} is used to upsample the feature maps for DM and SR. A detachable layer is attached to $\mathcal{F}_M$ to produce the intermediate output $M^{SR}$ for additional training supervision, and can be removed during testing. In our experiment, the number of RRDB modules for both $\mathcal{F}_M$ and $\mathcal{C}_M$ is set to $6$. 

\RR{Compared to E2ENet, TENet++ has two major differences: (1) the upsampling layer for SR is moved forward to the middle of the network (end of $\mathcal{F}_M$) to yield super-resolved raw images; (2) intermediate supervision is provided.
From a theoretical perspective, we hypothesize that reasonable intermediate supervision (superresolved raw image in our case) yields a limited solution space with good local minima, thus leading to an eased optimization.
We show the effectiveness of TENet++ over E2ENet through extensive experiments in Sec. \ref{sec:exp}.
} 

\begin{figure*}[ht]
\begin{center}
\includegraphics[width=1.0\linewidth]{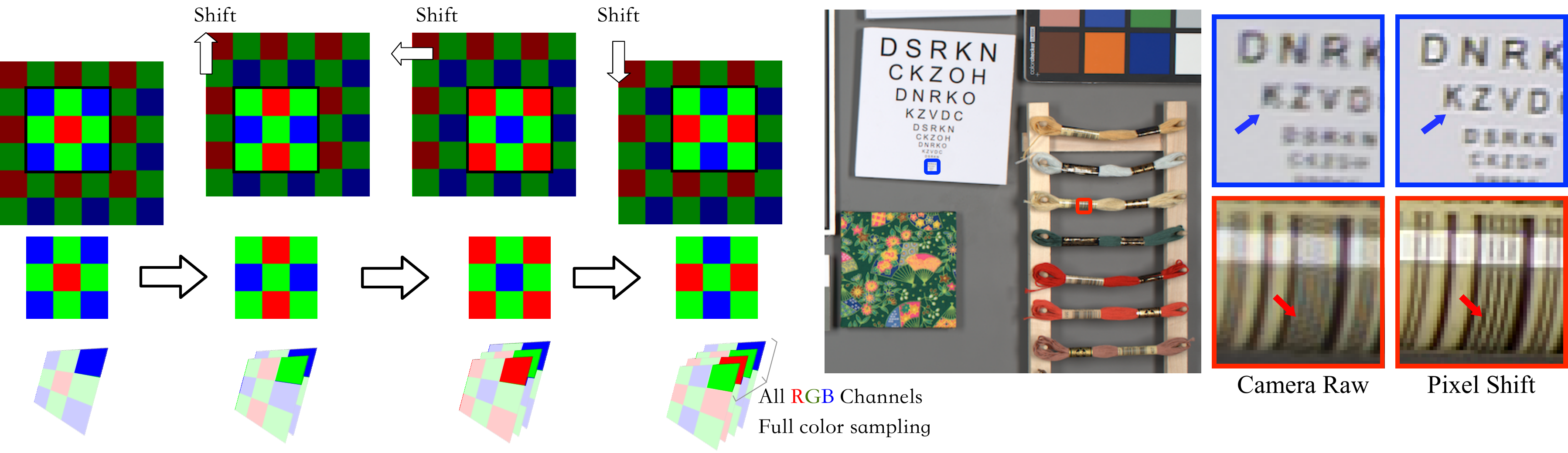}
\caption{
The pixel shift technique used to create dataset \textit{PixelShift200} (left) and qualitative comparison between the commonly used raw processing software Camera Raw, and the pixel shift. Pixel shift collects artifact-less (less zippering, moir\'e and chromatic aberration) full color sampled images directly without color interpolation. }
\label{fig:pixelshift}
\end{center}
\end{figure*}

\section{PixelShift200 Dataset}\label{sec:data}
\subsection{Motivation of PixelShift200}\label{sec:pixelshift200:motivation}
Previous learning-based DM algorithms train their networks on incompletely color-sampled datasets such as DIV2K \cite{DIV2K} and ImageNet \cite{imagenet}, where they take color images demosaiced from incomplete color samples (Bayer images) as Ground Truth and synthesize the mosaic images as input \cite{Brooks2019UnprocessingIF, Zamir2020CycleISPRI, Xing_2021_CVPR}. However, this scheme has three main issues: (1) the color images are interpolated by the camera ISP, which introduces DM artifacts caused by incomplete color sampling. These artifacts will also be learned if a DM model is trained on them. (2) The DM model trained on such synthesized dataset only learns an ``average'' DM algorithm used in the camera's ISP. And (3) the synthesized raw images only have a depth of 8-bit and therefore suffer from information loss, compared to normal 14-bit real raw images. Thus, real-world, high-resolution, uncompressed image datasets with full-color sampling are needed.

We contribute a novel real-world dataset \textit{PixelShift200}, which contains 200 4K-resolution full-color sampled images. The color information in red, green, and blue in 14-bit for each pixel is known in our dataset without any domosaicing. PixelShift200 was collected using the pixel shift technique \cite{sonypixelshift} embedded in the camera we use (see \secLabel \ref{exp:pixelshift200_collect}). This technique takes four samples of the same image at the same time, and physically controls the camera sensor to precisely move one pixel horizontally or vertically at each sampling. The four samples are then combined to directly obtain all the color information for each pixel. Refer to \figurename~\ref{fig:pixelshift} for an example of the pixel shift process. The pixel shift technique ensures that the sampled images follow the distribution of natural images.

Due to full-color sampling, our collected images in PixelShift200 are almost free of artifacts compared to the images interpolated from mosaic inputs. \figurename~\ref{fig:pixelshift} compares a color image obtained by the pixel shift technique with the output of the well-known raw processing software, Adobe Camera Raw (version 12.3). The pixel shift combines the four raw images into a single full-color sampled image, while Camera Raw interpolates the first sample using the built-in demosaicing algorithm.  It is worth noting that the pixel shift technique generates much less aliasing (see the letter ``K'' in the first row) and fewer moir\'es (see the barcode in the second row).
In \secLabel \ref{sec:ablation_data}, we demonstrate 
training raw image processing networks on our PixelShift200 dataset will produce better image quality than training the same network on the incompletely color sampled dataset (\eg DIV2K \cite{DIV2K}). 
We highlight that, as far as we are aware, we are the first to collect such a full-color sampled dataset. PixelShift200 is useful for training raw image processing methods and can also be used as a unique benchmark for demosaicing-related tasks. 

\subsection{PixelShift200 Collection Procedure}
\label{exp:pixelshift200_collect}
We collected PixelShift200 dataset with a Sony ILCE-7RM3 digital camera, which includes the pixel shift technique in its camera system \cite{sonypixelshift}. 
To avoid serious noise, we mounted a lens with fixed focal length and aperture (Zeiss FE $50$ $mm/1.4$) with low photosensitivity (ISO 100 or less). To reduce motion parallax, we controlled the depth of the scene field to a small range and held the camera with a heavy tripod. PixelShift200 consists of 200 4K resolution images for training and 20 1K resolution images for testing. \RR{The testing set is selected to cover a wide range of scenes.} As data augmentation, the training samples were cropped into $9444$ overlapping patches of size $512\times512$. 

\section{Experiments}\label{sec:exp}
\subsection{Experimental Setup}

\mysection{Data Preprocessing}
We perform a bicubic downsampling kernel (denoted as $\mathcal{S}^{-1}_{C}$), a mosaic kernel \cite{Brooks2019UnprocessingIF} ($\mathcal{C}^{-1}_{M}$), and then the Gaussian-Poisson noise model \cite{Hasinoff2014PhotonPN} to generate  LR noisy raw images $M_n^{LR}$ as input from HR color images $I^{HR}$ in pixelshift200: 
\begin{equation}\label{eqn:input_generate}
    M_n^{LR} = \mathcal{C}^{-1}_{M}(\mathcal{S}^{-1}_{C}(I^{HR}))+n
\end{equation}
where the noise term $n$ is sampled from:
\begin{equation}
    n \sim \mathcal{N}(\mu = 0, \sigma^2 = \lambda_{read}+\lambda_{shot}M^{LR})
\end{equation}
$\lambda_{read}$ and $\lambda_{shot}$ are the read and shot noise levels of a given raw image. The noise variance $n'$ is given by $(\lambda_{read}+\lambda_{shot}M_n^{LR})$.

In Pixelshift200, we generated the random Gaussian-Poisson noise in both training and testing. Note that the noise was generated on the fly during training and was sampled once and fixed for the testing samples. Noise levels follow the same range as the real-shot denoising benchmark dataset, DND \cite{Plotz2017BenchmarkingDA}.
Random rotation and flipping were used as data augmentation during training.
The output of the model after the whole pipeline is the RGB image in the linear color space. \RR{The black level subtraction is conducted as the pre-processing step for each raw image and is performed before DN, DM, and SR.} The white balance and color mappings were read from the raw images and applied to the final outputs to transform them into standard RGB space (sRGB).

\mysection{Metric}
\RR{For quantitative experiments, we use PSNR ($\uparrow$), SSIM ($\uparrow$), and FreqGain ($\downarrow$) \cite{gharbi2016deep} to measure overall fidelity, overall structure similarity, and fine-grained artifacts. 
Note that FreqGain is the metric we modify from \cite{gharbi2016deep}, which was proposed to detect moir\'es. We revise it to a scalar version by averaging the positive logarithmic values of the frequency gains. The formula of FreqGain is the following:
\begin{equation}
    \rho=\text{avg}\left(\text{ReLU}\left(\log \left(\frac{ \vert \mathcal{F_O(\omega)}\vert^2+\epsilon}{\vert \mathcal{F_I(\omega)}\vert^2+\epsilon}\right) \right)\right)
\end{equation}
where $\mathcal{F_I(\omega)}$ and $\mathcal{F_O(\omega)}$ represent the 2D Fourier transform of the ground truth and the prediction. $\epsilon=10^{-6}$ is added to avoid dividing by zero. ReLU is used to only consider positive values that represent regions where moir\'e-like artifacts are likely to appear.  Averaged value across frequencies is returned as the quantitative metric. 
}

\mysection{Network Training}
We optimized all models using Adam \cite{Adam} with an initial learning rate $lr=5\times10^{-4}$ on four NVIDIA RTX2080Ti GPU. A cosine annealing learning rate schedule is adopted. \RR{All models are trained for 1000 epochs to ensure convergence.} 

\mysection{Experimental Setup of Comparison with the State-of-the-art}
The most closely related works are JDSR \cite{zhou2018deep},  RawSR \cite{Xu2019TowardsRS},  SGNet \cite{Liu2020JointDA}, and JDnDmSR \cite{XingEndtoEndLF}, where most of which are black-box end-to-end networks without a specific pipeline (\textbf{E2ENet}). Since the architectures and data processing are different, rather than unfairly comparing with these networks, we implemented all possible pipelines (including E2ENet) using the same module as TENet++ (see \figLabel \ref{fig:arch_e2e} for the module structure) and trained all networks on the same PixelShift200 dataset. We also validate our proposed pipeline on different datasets and using models built by different modules.

\begin{figure*}[ht]
\begin{center}
\includegraphics[width=1.0\linewidth]{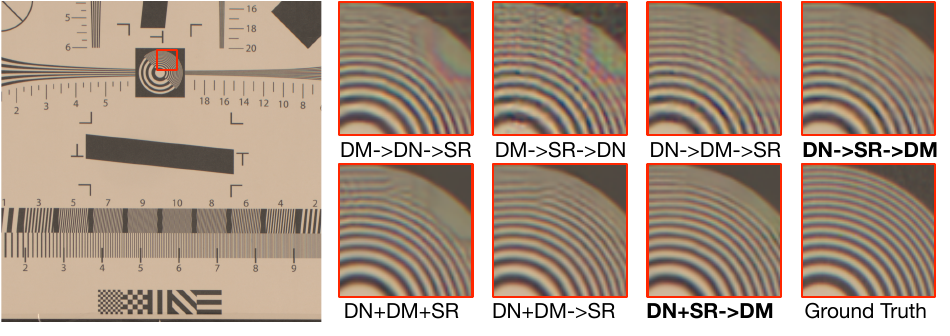}
\caption{
\textbf{Qualitative comparisons of different pipelines on an example from PixelShift200 test set.} The left is the ground truth image, while the right shows the closeups of the output of different pipelines. The input is the low-resolution noisy mosaiced version of the left image. The top row of the right shows the results using different sequential solutions, while the bottom row shows the results of fully joint pipelines and the ground truth. Our proposed pipeline \pipeline yields the highest quality among all sequential pipelines, while \pipelinejoint achieves the best among all the joint pipelines.
}
\label{fig:result_pipeline}
\end{center}
\end{figure*}

\begin{figure*}[ht]
\begin{center}
\includegraphics[width=1.0\linewidth]{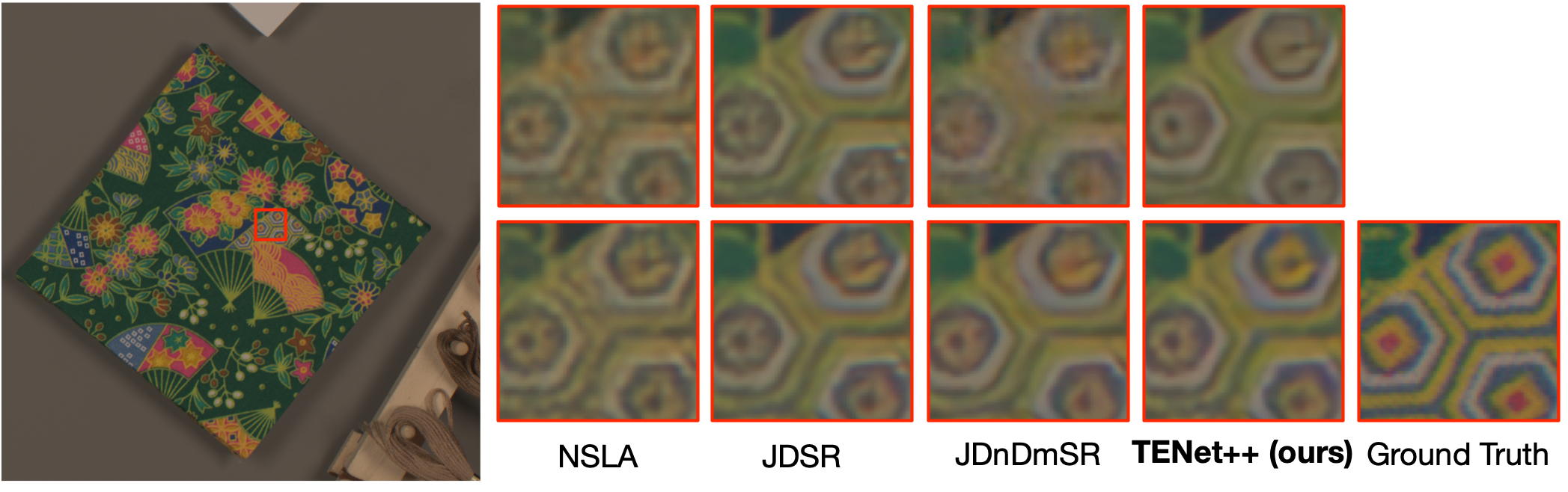}
\caption{
\textbf{Qualitative comparisons between DN$+$DM$+$SR (top row) \vs \pipelinejoint~(bottom row) on different architectures on PixelShift200 test set.}
Our pipeline produces results with sharper edges and preserves the color of the objects better.
}
\label{fig:pipeline_architecture}
\end{center}
\end{figure*}

We compare our proposed pipeline \pipeline with all other possible pipelines in sequential solutions, and our \pipelinejoint pipeline with others in partially and fully joint solutions. A sequential solution applies three separate models sequentially, \eg DN$\to$SR$\to$DM executing DN, SR, and DM sequentially. A partially joint solution sequentially conducts two models, while one is a joint model for two tasks, and another a single-task model. For example, DN+SR$\to$DM performs first a joint DN and SR model DN+SR, and then a DM model.
A fully joint solution solves the three tasks together using a single model. The pipeline of a joint solution (\eg, DN+SR$\to$DM) is achieved by providing additional supervision (\eg, denoised super-resolved mosaic image) in an end-to-end network. 
The way of providing intermediate supervision is mentioned in \secLabel \ref{sec:pip_joint}. 
\RR{
All models needed are implmented as follows:
\begin{itemize}
    \item Five single-task models: raw image denoising, raw image SR, demosaicing, color image denoising, and color image SR. All five models are implemented in the same way as $\mathcal{F}_M$ (\figLabel \ref{fig:net}) by 6 RRDBs. 
    \item Three partially joint models: DN+DM (joint DN and DM), DN+SR (joint raw image DN and SR), and DM+SR (joint DM and SR). While DN+DM and DN+SR are implemented as $\mathcal{F}_M$ (\figLabel \ref{fig:net}) using 6 RRDBs, DM+SR are implemented as E2ENet using 12 RRDBs.
    \item Four fully joint models: DN+SR$\to$DM (proposed TENet++), DN+DM+SR (E2ENet), DN$\to$DM+SR (similar architecture as TENet++ where DN supervision is provided instead), and DN+DM$\to$SR (similar architecture as TENet++ where DM supervision is provided instead). All fully joint models are implemented by 12 RRDBs.
\end{itemize}
}

All models are trained with the $\times 2$ SR factor and the same level of Gaussian-Poisson noise in PixelShift200. We compare our proposed pipelines with other pipelines using these models for a fair comparison.

\subsection{Pipeline Comparison Experiments}\label{sec:exp_pip}
\begin{table}[t]
\footnotesize
\begin{center}
\caption{\textbf{Comparison of pipelines on PixelShift200 test set}. Gaussian-Poisson noise with $\times2$ SR are used. \textbf{Bold} denotes the best performance. Our proposed pipelines yield the best quantitative results among all possible pipelines.}
\label{tab:pipeline}
\resizebox{\columnwidth}{!}{
\begin{tabular}{ll|ccc}
\toprule
\textbf{Type}&
\textbf{Pipeline}&
PSNR & SSIM & FreqGain$\downarrow$\\
\midrule
\multirow{6}{*}{Sequential} &
DM$\to$DN$\to$SR (usual) &
33.51 & 0.8379 & \RR{0.4853}\\
&DM$\to$SR$\to$DN & 
30.01 & 0.6773 &\RR{1.0978} \\&
SR$\to$DM$\to$DN &
31.44 & 0.7270 &\RR{0.7858} \\&
SR$\to$DN$\to$DM &
33.42 & 0.8059 &\RR{0.5583}\\&
DN$\to$DM$\to$SR &
36.33 & 0.9256 &\RR{0.2067}\\\rowcolor{gray!40} \cellcolor[gray]{1.0}&
\textbf{DN$\to$SR$\to$DM \textbf{(ours)}} &
 \textbf{36.61} & \textbf{0.9294} &\RR{\textbf{0.1886}}\\
\midrule

\multirow{3}{*}{Partially joint} &
DN$\to$DM+SR &
36.65 & 0.9299 &\RR{0.1884}\\ &
DN+DM$\to$SR  &
36.24 & 0.9259 &\RR{0.1952}\\
\rowcolor{gray!40} \cellcolor[gray]{1.0}&
 \textbf{DN+SR$\to$DM} \textbf{(ours)}  &
\textbf{37.04} & \textbf{0.9327} &\RR{\textbf{0.1829}}\\

\midrule
\multirow{3}{*}{Fully joint} &
DN+DM+SR &
36.71 & 0.9292 &\RR{0.1851}\\ &
DN$\to$DM+SR  &
{36.18} &	{0.9245} &\RR{0.2451}\\&
DN+DM$\to$SR  &
{37.24} &	{0.9341} &\RR{0.1907}\\\rowcolor{gray!40} \cellcolor[gray]{1.0}&
\textbf{DN+SR$\to$DM (ours)} &
\textbf{37.36} & \textbf{0.9353} &\RR{\textbf{0.1814}}\\

\bottomrule
\end{tabular}}
\end{center}
\end{table}
\mysection{Proposed pipeline outperforms others in sequential solutions.}
\tablename~\ref{tab:pipeline} shows that \textit{our proposed pipeline \pipeline clearly outperforms all other pipelines under the sequential solution setting.}
Surprisingly, the PSNR is 3.10 dB higher for our pipeline than the usual pipeline DM$\to$DN$\to$SR. This improvement is achieved simply by adopting our pipeline as a replacement for the other pipelines.
As observed, when DN is not performed in the first stage, the image quality obtained will drop sharply. When DN is fixed as the first task, our proposed pipeline still improves PSNR by 0.28 dB, which reflects that performing SR before DM yields a higher quality than performing DM before SR. These experimental findings confirm our discussion in \secLabel \ref{sec:new_pipeline} \RR{that DN and SR in mosaic space are suggested}.

\mysection{Proposed pipeline \RR{insignificantly} outperforms others in joint solutions.}
Since performing DN in the first stage is the best option, we now mainly study the pipelines where DN is performed first among partially and fully joint solutions. 
\tablename~\ref{tab:pipeline} shows the quantitative comparisons. 
One can conclude that (1) joint solutions outperform sequential solutions in the same execution order, as expected. For example, DN+SR$\to$DM in both partially and fully joint solutions produces images with better metric values than DN$\to$SR$\to$DM in sequential solutions. 
(2) \textit{In both partially joint and fully joint solutions, our proposed pipeline \pipelinejoint~consistently generates slightly higher PSNR and SSIM than other pipelines}. 
In particular, PSNR of the proposed pipeline is 0.39 dB higher than any other pipelines in the partially joint solutions. In joint solutions, TENet++ (\pipelinejoint) outperforms the E2ENet counterpart (DN+DM+SR) by 0.65 dB in terms of PSNR.
\RR{However, we highlight that the improvement of the proposed pipeline in joint solutions is less than 1 dB, which is not as significant as sequential solutions. Such a marginal improvement may indicate that the execution order of tasks in an end-to-end solution might be inapplicable.
}

\mysection{Qualitative comparisons of different pipelines.}
The comparison of sequential solutions in \figLabel \ref{fig:result_pipeline} (top row) shows our proposed pipeline \pipeline~clearly outperforms other pipelines with 
significantly fewer color artifacts, validating our suggestion to perform SR before DM. Our pipeline produces less noise and reflects the importance of performing DN at the first stage. 
In the fully joint solutions (\figLabel \ref{fig:result_pipeline} bottom row), our proposed pipeline again achieves better qualitative results than others, suffering less moir\'e. 
\section{Ablation Study} \label{sec:ablation}

\begin{table}[t]
\footnotesize
\begin{center}
\caption{\textbf{Ablation on architectures.}
We experiment with the other possible architectures constructed by the NLSA \cite{Mei_2021_CVPR} block and two SOTA models, JDSR \cite{zhou2018deep} and JDnDmSR \cite{XingEndtoEndLF}. 
Our proposed pipeline \pipelinejoint consistently improves the performance of all given networks on the task of joint DN, DM and SR.
}
\label{tab:ablation_archi}
\resizebox{\columnwidth}{!}{
\begin{tabular}{llcccc}
\toprule
\textbf{Architecture} & \textbf{Pipeline} & PSNR & SSIM & FreqGain$\downarrow$\\
\midrule
\multirow{2}{*}{NLSA block \cite{Mei_2021_CVPR}} &
DN+DM+SR &
34.63 & 0.9086 & 0.2975\\ &
\textbf{DN+SR$\to$DM (ours)} &
\textbf{36.05} & \textbf{0.9270} & \textbf{0.1769}\\
\midrule
\multirow{2}{*}{JDSR \cite{zhou2018deep}} &
DN+DM+SR &
36.53 & 0.9289 & 0.1957 \\ &
\textbf{DN+SR$\to$DM (ours)} &
\textbf{36.68} & \textbf{0.9296} &\textbf{0.1947}\\
\midrule
\multirow{2}{*}{JDnDmSR \cite{XingEndtoEndLF}} &
DN+DM+SR & 33.11 & 0.8782 & 0.4180 \\
& \textbf{DN+SR$\to$DM (ours)} &\textbf{36.91} & \textbf{0.9317} & \textbf{0.1959} \\ 
\midrule
\multirow{2}{*}{TENet++ (Ours)} &
DN+DM+SR & 36.71 & 0.9292 & 0.1851 \\
& \textbf{DN+SR$\to$DM (ours)} &\textbf{37.36} & \textbf{0.9353} &\RR{\textbf{0.1814}} \\ 

\bottomrule
\end{tabular}}
\end{center}
\end{table}
\begin{figure*}[!htb]
\begin{center}
\includegraphics[width=\linewidth]{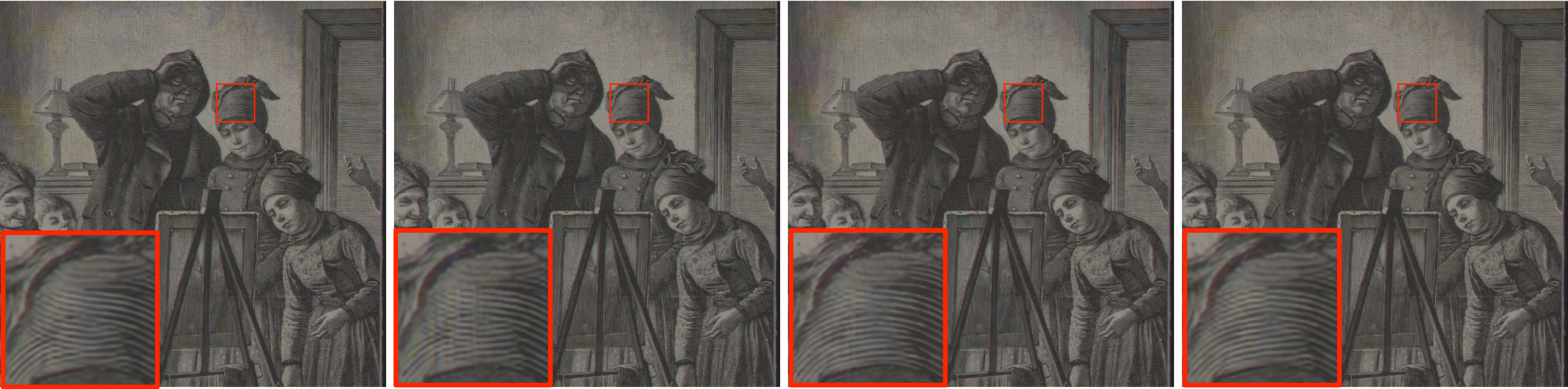}
\vspace{2em}
{\parbox{\linewidth}{ \vspace{3mm}\begin{overpic}[width=\linewidth]{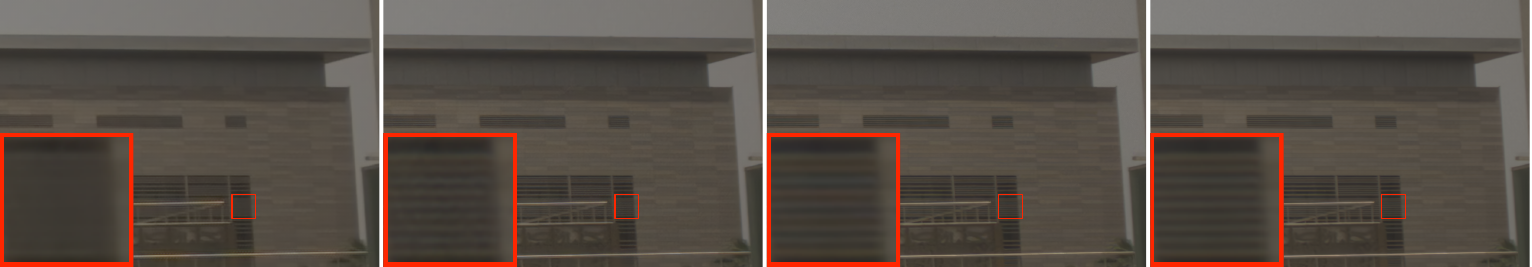}
\put(8,-4){\scalebox{1.0}{JDSR\cite{zhou2018deep}}}
\put(32,-4){\scalebox{1.0}{JDnDmSR \cite{XingEndtoEndLF}}}
\put(56,-4){\scalebox{1.0}{TENet++ (DIV2K)}}	\put(79,-4){\scalebox{1.0}{TENet++ (PixelShift200)}}
\end{overpic}}}
\caption{
\textbf{Qualitative comparisons on the real-shot images}. We compare the SOTA methods JDSR \cite{zhou2018deep}, JDnDmSR \cite{XingEndtoEndLF} and our proposed TENet++ trained on DIV2K, and TENet++ trained on 
PixelShift200. Images were captured using a Sony ILCE-7RM3 (top row) and iPhone XS Max (bottom row).}
\label{fig:ablation_dataset_realshot}
\end{center}
\end{figure*}
\begin{figure*}[!htb]
\begin{center}

{\parbox{1.0\textwidth}{\begin{overpic}[width=\linewidth]{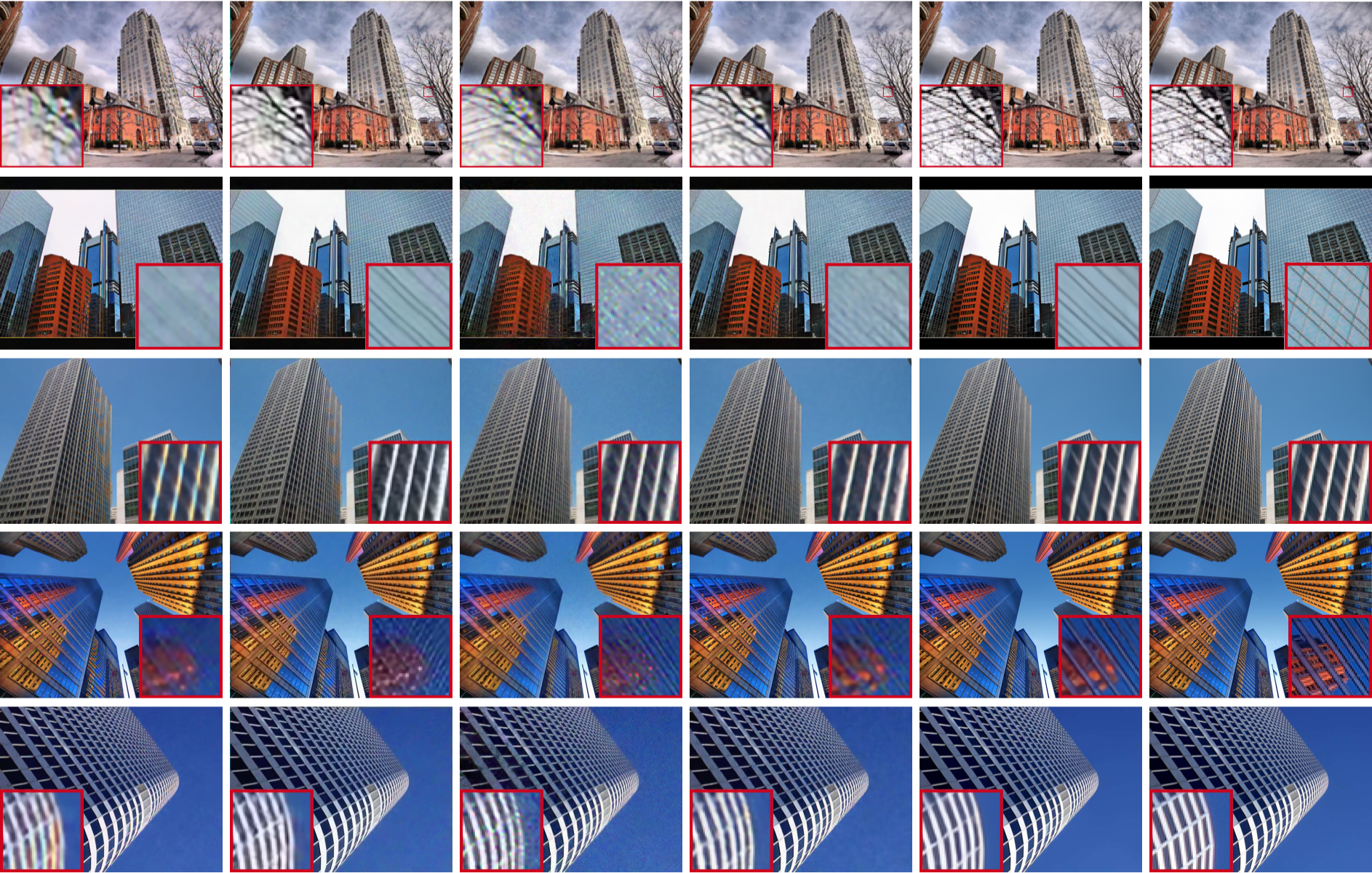}
\put(3,-3){\scalebox{1.0}{ADMM \cite{tan2017joint}}}
\put(20,-3){\scalebox{1.0}{Condat \cite{condat}}}
\put(36,-3){\scalebox{1.0}{FlexISP \cite{flexisp}}}
\put(51,-3){\scalebox{1.0}{DemosaicNet\cite{gharbi2016deep}}}
\put(69,-3){\scalebox{1.0}{TENet++ (ours)}}
\put(86,-3){\scalebox{1.0}{Ground Truth}}
\end{overpic}
\vspace{1mm}
}}
\end{center}
\caption{\textbf{The qualitative comparison of different methods on the noisy Urban100 \cite{Urban100} test images.} The noise model is the additive Gaussian noise (sigma=10) and the SR factor is 2. Our TENet achieves a close performance to the Ground Truth.}
\label{fig:noisy}
\end{figure*}

\begin{table*}[t]
\footnotesize
\begin{center}
\caption{\textbf{Ablation on datasets.}
Models are trained on DIV2K \cite{DIV2K}, and tested on DIV2K test set, Urban100 \cite{Urban100}, and CBSD68 \cite{BSD100} with Gaussian-Poisson noise and $\times2$ SR. Our proposed pipeline outperforms other pipelines \RR{in both sequential and joint solutions} regardless of datasets.}
\label{tab:ablation_dataset}
\resizebox{2\columnwidth}{!}{
\begin{tabular}{ll|ccc|ccc|ccc}
\toprule
\multirow{2}{*}{\textbf{Type}} & \multirow{2}{*}{\textbf{Pipeline}} & \multicolumn{3}{c|}{DIV2K \cite{DIV2K}} & \multicolumn{3}{c|}{Urban100 \cite{Urban100}} & \multicolumn{3}{c}{CBSD68 \cite{BSD100}}\\ \cmidrule{3-11}
&&
PSNR & SSIM & FreqGain$\downarrow$&
PSNR & SSIM & FreqGain$\downarrow$&
PSNR & SSIM & FreqGain$\downarrow$\\
\midrule
\multirow{6}{*}{\RR{Sequential}} &
 DM$\to$DN$\to$SR (usual) & 20.02 & 0.6296	& 0.8705	& 19.32 &	0.6393&	1.7710	&20.95&	0.6399 &	0.8073\\
 & DM$\to$SR$\to$DN & 19.60	& 0.5684&	1.1938&	18.96	&0.5838&	1.9813&	20.55&	0.5989	&1.0690\\
 &SR$\to$DM$\to$DN & 19.88&	0.5950&	0.7514	&19.25&	0.6103&	1.5520&	20.83	&0.6195&	0.6139\\
 & SR$\to$DN$\to$DM & 20.07 &	0.6309&	0.6012&	19.41	&0.6379&	1.4749&	21.00	&0.6446&	0.4723\\
 &DN$\to$DM$\to$SR &20.30&	\textbf{0.7660}&	0.4046&	19.52&	\textbf{0.7326}&	1.5211&	21.15&	\textbf{0.7097}&	0.4599\\
 \rowcolor{gray!40} \cellcolor[gray]{1.0}&
 \textbf{DN$\to$SR$\to$DM \textbf{(ours)}} &
\textbf{20.30} &	0.7602&	\textbf{0.2748}	&\textbf{19.55}&0.7282&	\textbf{1.3483}&	\textbf{21.17}	&0.7079&	\textbf{0.2837}\\ \midrule
\multirow{3}{*}{\RR{Partially joint}} &
 DN$\to$DM+SR & 20.30 &	0.7617&	0.2875&	19.59&	0.7337	&1.3684	&21.15&	0.7040 &	0.3260\\
 & DN+DM$\to$SR  &20.30	& \textbf{0.7664} &	0.4217 &	19.55	& 0.7366 &	1.5405 &	21.15	& 0.7101 &	0.4661 \\
 \rowcolor{gray!40} \cellcolor[gray]{1.0}&
 \textbf{DN+SR$\to$DM} \textbf{(ours)} & \textbf{20.34} &	0.7630&	\textbf{0.2693}&	\textbf{19.67}&	\textbf{0.7391}&	\textbf{1.3379}&	\textbf{21.22}	&\textbf{0.7109}&	\textbf{0.2776}\\\midrule
\multirow{2}{*}{Fully joint} &DN+DM+SR &20.30 &	0.7563&	0.3167&	19.59&	0.7295&	1.3980&	21.16&	0.7019&	0.3571\\
&
\cellcolor[gray]{0.8}{\textbf{DN+SR$\to$DM (ours)}}
&\cellcolor[gray]{0.8}\textbf{20.37}
&\cellcolor[gray]{0.8}\textbf{0.7677}
&\cellcolor[gray]{0.8}\textbf{0.2766}
&\cellcolor[gray]{0.8}\textbf{19.72}
&\cellcolor[gray]{0.8}\textbf{0.7467}
&\cellcolor[gray]{0.8}\textbf{1.2965}
&\cellcolor[gray]{0.8}\textbf{21.24}
&\cellcolor[gray]{0.8}\textbf{0.7118}
&\cellcolor[gray]{0.8}\textbf{0.3216}
\\
\bottomrule
\end{tabular}}
\end{center}
\end{table*}


\begin{table}[hbt]
\footnotesize
\begin{center}
\caption{\textbf{Ablation on SR factor.} SR factor is 4. The proposed pipeline again outperforms other pipelines in both sequential and partially joint solutions.}
\label{tab:pipeline_sr}
\resizebox{\columnwidth}{!}{
\begin{tabular}{ll|ccc}
\toprule
\textbf{Type}&
\textbf{Pipeline}&
PSNR & SSIM & FreqGain$\downarrow$\\
\midrule
\multirow{6}{*}{Sequential} &
DM$\to$DN$\to$SR (usual) &
31.49 & 0.8347 &0.2473\\&
DM$\to$SR$\to$DN &
28.05 & 0.6766 &0.7773\\&
SR$\to$DM$\to$DN &
28.70&0.6561 &0.5657\\&
SR$\to$DN$\to$DM &
29.25&0.6752 &0.5200\\&
DN$\to$DM$\to$SR &
32.41&0.8654 &0.1650\\\rowcolor{gray!40} \cellcolor[gray]{1.0}&
\textbf{DN$\to$SR$\to$DM \textbf{(ours)}} &
\textbf{32.99}	& \textbf{0.8752} &\textbf{0.1506}\\
\midrule

\multirow{3}{*}{Partially joint} &
DN$\to$DM+SR &
32.95 &  0.8739 &0.1593\\ &
DN+DM$\to$SR  &
32.32 & 0.8665 & \textbf{0.1542}\\\rowcolor{gray!40} \cellcolor[gray]{1.0}&
 \textbf{DN+SR$\to$DM} \textbf{(ours)}  &
\textbf{33.06} & \textbf{0.8770} &0.1547\\
\midrule
\multirow{4}{*}{Fully joint} &
DN+DM+SR &
33.48 & 0.8797 &0.1656 \\ &
DN$\to$DM+SR  &
33.21 & 0.8765 &0.1917\\&
DN+DM$\to$SR  &
33.45 & 0.8796 &0.1893 \\ 
\rowcolor{gray!40}
\cellcolor[gray]{1.0}
&
\textbf{DN+SR$\to$DM (ours)} &
\textbf{33.54} & \textbf{0.8810} &\textbf{0.1655}\\
\bottomrule
\end{tabular}}
\end{center}
\end{table}

\begin{table}[t]
\footnotesize
\begin{center}
\caption{\textbf{Ablation on noise model.} We experiment a different noise model, the additive Gaussian noise (sigma 10), with $\times 2$ SR. Models are trained on DF2K \cite{lim2017enhanced}. \RR{Our proposed pipeline \pipelinejoint outperforms DN+DM+SR.}}
\label{tab:pipeline_gaussian}
\resizebox{\columnwidth}{!}{
\begin{tabular}{llccc}
\toprule
\textbf{Dataset} & \textbf{Pipeline} & PSNR & SSIM & FreqGain$\downarrow$\\\toprule
\multirow{2}{*}{\RR{DIV2K} \cite{DIV2K}} & DN+DM+SR &29.74 &	0.8396 & 0.3113\\

&\textbf{DN+SR$\to$DM (ours)} & \textbf{29.81} &	\textbf{0.8410} &	\textbf{0.2978}\\\midrule
\multirow{2}{*}{Urban100 \cite{Urban100}} & DN+DM+SR &26.81 &	0.8287 &	1.2128\\
&\textbf{DN+SR$\to$DM (ours)} & \textbf{26.96}	& \textbf{0.8327} &	\textbf{1.1960} \\\midrule
\multirow{2}{*}{\RR{CBSD68} \cite{BSD100}} & DN+DM+SR &27.52 &	0.7766 &	0.3674\\
&\textbf{DN+SR$\to$DM (ours)} & \textbf{27.56} &	\textbf{0.7775} &	\textbf{0.3616}\\
\bottomrule
\end{tabular}}
\end{center}
\end{table}

\subsection{Ablate Proposed Pipeline}
We have demonstrated that our proposed pipeline is quantitatively and qualitatively better than other pipelines in \secLabel \ref{sec:exp}. However, one may wonder: (1) what if a different architecture is used other than the RRDB module and TENet++? (2) What if a different dataset instead of Pixelshift200 is used for training and evaluation? (3) What if a different SR factor is used instead of 2? (4) What if a different noise model is adopted instead of the Gaussian-Poisson noise model?
Here \textit{we validate that our proposed pipelines consistently outperform other pipelines in a variety of settings}.

\mysection{Architecture.} 
We ablate the module-level and network-level architectures in PixelShift200. 
\RR{The module-level architecture ablation study denotes that we use the same architecture as E2ENet for pipeline DN+DM+SR and as TENet++ for pipeline \pipelinejoint where a different module (\eg NLSA) is used instead of the original RRGB. The network-level architecture ablation study means a different architecture rather than TENet++ is used.}
For the module-level experiment, we leverage the non-local sparse attention (NLSA) module from the state-of-the-art (SOTA) image SR work \cite{Mei_2021_CVPR} to build the end-to-end network. For the network-level experiment, we replace TENet++ with the SOTA networks JDSR \cite{zhou2018deep} and JDnDmSR \cite{XingEndtoEndLF} for the joint DN, DM and SR problem. \RR{We insert our proposed pipeline into the two models by providing intermediate supervision in a similar way as TENet++ as illustrated in \figLabel \ref{fig:arch_tenet}.} \tablename~\ref{tab:ablation_archi} compares the performance of the original pipeline (DN+DM+SR) and the same network using our proposed pipeline \pipelinejoint. Experiments on all three architectures show 
\textit{our pipeline consistently improves the performance regardless of the architecture designs}. 
\RR{In addition, by comparing results in Tab. \ref{tab:ablation_archi} with Tab. \ref{tab:pipeline}, one can observe that our proposed TENet++ outperforms the network constructed by the SOTA module and the SOTA networks (JDSR \cite{zhou2018deep}, JDnDmSR \cite{XingEndtoEndLF}) for joint DN, DM, and SR.}
Qualitative comparisons of the results of different architectures (columns) fitted with two different pipelines (the top row shows the usual pipeline DN+DM+SR, the bottom row shows our pipeline \pipelinejoint) are presented in \figLabel \ref{fig:pipeline_architecture}. For each network, our pipeline \pipelinejoint in joint solution enhances image sharpness while also preserving the color of the objects to a greater extent. \RR{Our TENet++ also yields more visually appealing images than SOTA when equipped with the same pipeline.}

\mysection{Dataset.}
We also experiment with different pipelines (\RR{sequential and joint solutions}) on other datasets instead of PixelShift200. We train models with different pipelines on DIV2K 800 training images \cite{DIV2K}, where the mosaic images are synthesized from color images using the same unprocessing technique in \cite{Brooks2019UnprocessingIF}.
Gaussian-Poisson noise and $\times2$ SR are used.
The evaluation on three widely used benchmarks, the DIV2K test set, Urban100 \cite{Urban100}, and CBSD68 \cite{BSD100} is provided in \tablename~\ref{tab:ablation_dataset}. As observed, \textit{our proposed pipelines improves the network's performance across all the widely-used benchmarks in both sequential and joint solutions}.
\RR{Despite the consistent improvement, the PSNR gain in joint solution is less than 0.2 dB in all benchmarks, which again shows that shuffling the pipeline in an end-to-end network is not necessarily applicable.}

\mysection{SR factor.}
We also experiment with a different factor ($\times 4$) of super-resolution to validate the benefit of the proposed pipeline. \tablename~\ref{tab:pipeline_sr} shows that \textit{our proposed pipeline outperforms other pipelines under the $\times 4$ SR factor} in both sequential and joint solutions.

\mysection{Noise model.}
Our previous experiments are conducted under the Gaussian-Poisson noise modeling assumption. Here, we further validate our pipeline under a different assumption of the noise model. We study the widely used Gaussian noise. The noise level (sigma) is set to 10. We train models on DF2K \cite{lim2017enhanced} \RR{and evaluate on the DIV2K test set \cite{DIV2K}, Urban100 \cite{Urban100} and CBSD68 \cite{BSD100}}. \textit{\tablename~\ref{tab:pipeline_gaussian} shows that our proposed pipeline \pipelinejoint is only able to marginally outperform the vanilla pipeline DN+DM+SR under the Gaussian noise setting} \RR{in joint solutions}. In \figurename \ref{fig:noisy}, we further show the qualitative results of our TENet++ compared to the previous methods with a pipeline of DN+DM$\to$SR. It is worth noting that our method achieves the closest qualitative performance to the Ground Truth.

\subsection{Ablate proposed Dataset PixelShift200}\label{sec:ablation_data}
We evaluate two identical models (TENet++) trained on two distinct datasets, our PixelShift200 and the incompletely color-sampled dataset DIV2K \cite{DIV2K}. The real-shot raw images are used as input. \textit{PixelShift200 helps the model suffer less 
moir\'e and color artifacts}, as shown in \figLabel \ref{fig:ablation_dataset_realshot} (column 3 \vs column 4). The improved qualitative performance is attributed to the full color-sampling and natural image distribution characteristics of the proposed PixelShift200.

\section{Real-Shot Experiments}
We compare TENet++ with the raw image processing library, DCRaw, and 
popular commercial software, Camera Raw, on a raw image shot with an iPhone X (see \figLabel~\ref{fig:teaser}).
\RR{The SR model implemented using the \figLabel \ref{fig:arch_e2e} network structure by 6 RRDBs (refer to \secLabel \ref{sec:exp_pip} for details) is used to super-resolve the demoisaiced outputs of DCRaw and Camera Raw.}
The proposed TENet++ yields clean results with rich detail. 
We also provide more real-shot comparisons between and our TENet++ and SOTA methods when equipped with the same pipeline (\pipelinejoint) as TENet++ in \figLabel \ref{fig:ablation_dataset_realshot}. 
All models are trained on DIV2K for a fair comparison. 
Compared to JDSR \cite{zhou2018deep}, our TENet++ successfully reconstructs the high-frequency texture. TENet++ also produces far fewer artifacts, such as moir\'es (refer to the scarf texture in the top row) and color aliasing (refer to the steel railing in the bottom row), than JDnDmSR \cite{XingEndtoEndLF}.


\section{Conclusion}
\RR{We presented intermediate supervision that enforces a certain pipeline in an end-to-end network.
We performed \RR{a comprehensive study in the effect of pipelines on the task of learning-based} denoising (DN), demosaicing (DM), and super-resolution (SR) in both sequential and joint solutions.
We found that the effect of the pipeline is significant in sequential solutions, while it is marginal in joint solutions, and thus shuffling the execution order of tasks is not very necessary for an end-to-end network. 
We also contributed PixelShift200, a full-color sampled dataset, for training and evaluating raw image processing-related tasks.}

\section{Limitation and Future work}
\RR{
First, the proposed PixelShift200 only includes static objects and has a limited size (200 unique samples). It will be more beneficial to the community if more samples could be collected. 
Second, this work only considers single-frame image processing. With increasing interest in the use of multiple frames \cite{wronski2019handheld}, we believe that it is promising to study end-to-end networks for multi-frame DN, DM, and SR, which have greater practical values but are rather under-explored.
}

\mysection{\RR{Acknowledgement}}\label{sec:ack}
\RR{The authors thank the reviewers of ICCP 2022 for valuable suggestions and Dr. Silvio Giancola for proofreading the rebuttals.
This work was supported by the KAUST Office of Sponsored Research (OSR) through the Visual Computing Center (VCC) funding.}

\bibliographystyle{IEEEtran}
\bibliography{main}

\ifpeerreview \else


\vspace{-2em}
\begin{IEEEbiography}[{\includegraphics[width=1in,height=1.25in,clip,keepaspectratio]{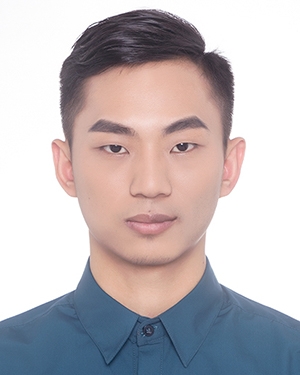}}]{Guocheng Qian}
is currently working towards a doctoral degree in the Department of Computer Science at King Abdullah University of Science and Technology (KAUST). He received his Master's degree from KAUST in 2020 and his BEng degree with first-class honors from Xi'an Jiaotong University (XJTU) in 2018. His research interests are in computer vision and geometric deep learning. He has published related papers in top conferences and journals, including CVPR, NeurIPS, T-PAMI, \etc
\end{IEEEbiography}

\vspace{-1em}
\begin{IEEEbiography}[{\includegraphics[width=1in,height=1.25in,clip,keepaspectratio]{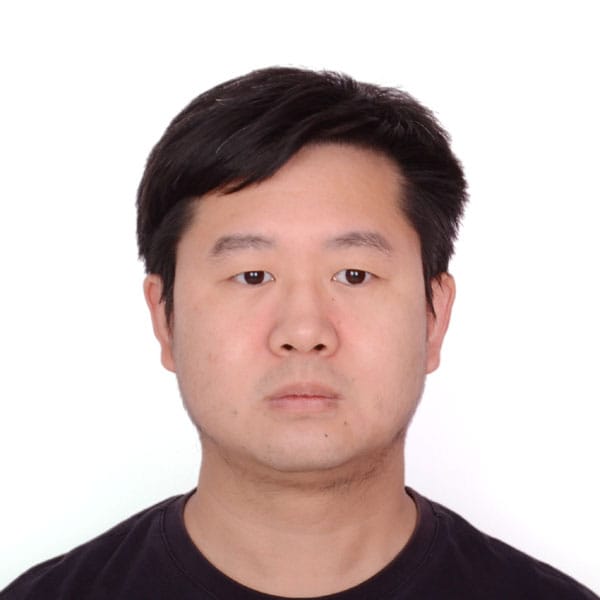}}]{Yuanhao Wang}
received the BEng degree from Beijing University of Posts and Telecommunications in 2013, and MEng degree from Tsinghua University in 2016. He is working towards the doctoral degree currently in the department of Electrical and Computer Engineering at King Abdullah University of Science and Technology. His research interests line in compuational imaging and neural radiance field. 
\end{IEEEbiography}

\vspace{-1em}
\begin{IEEEbiography}[{\includegraphics[width=1in,height=1.25in,clip,keepaspectratio]{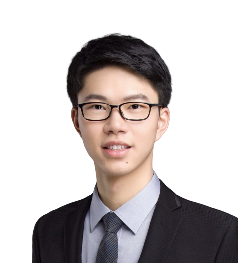}}]{Jinjin Gu}
is currently pursuing a Ph.D. degree in Engineering and IT with the University of Sydney. He received his B.Eng. degree in computer science and engineering from the Chinese University of Hong Kong, Shenzhen, in 2020. His research interests include computer vision, image processing, interpretability of deep learning algorithms, and machine learning applications in industrial. 
\end{IEEEbiography}

\vspace{-1em}
\begin{IEEEbiography}[{\includegraphics[width=1in,height=1.25in,clip,keepaspectratio]{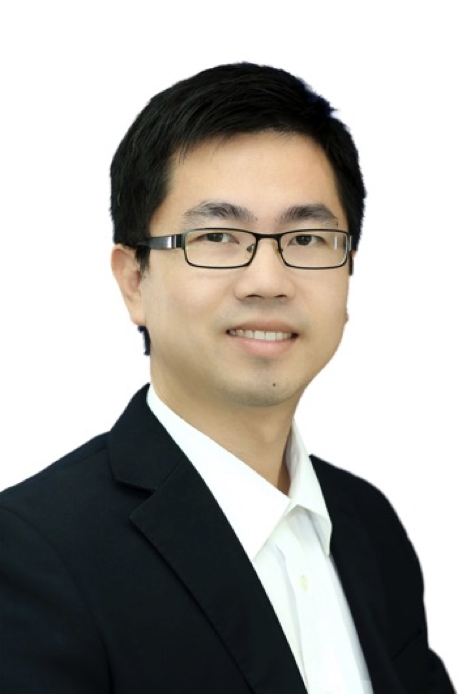}}]{Chao Dong}
is currently an associate professor at Shenzhen Institute of Advanced Technology, Chinese Academy of Science. He received his Ph.D. degree from The Chinese University of Hong Kong in 2016. In 2014, he introduced the deep learning method -- SRCNN into the super-resolution field. This seminal work was chosen as one of the top ten “Most Popular Articles” of TPAMI in 2016. His team has won several championships in international challenges -- NTIRE2018, PIRM2018, NTIRE2019, NTIRE2020 and AIM2020. He worked in SenseTime from 2016 to 2018 as the team leader of Super-Resolution Group. His Google citation has surpassed 16,000. His current research interest focuses on low-level vision problems, such as image/video super-resolution, denoising and enhancement. Email: \texttt{chao.dong@siat.ac.cn}. 
\end{IEEEbiography}

\begin{IEEEbiography}[{\includegraphics[width=1in,height=1.25in,clip,keepaspectratio]{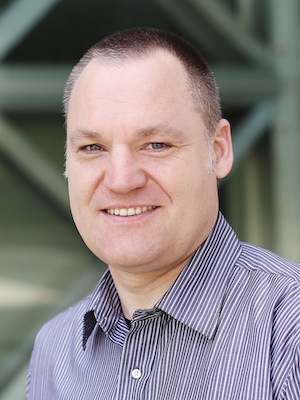}}]{Wolfgang Heidrich} (Fellow, IEEE) is a Professor of Computer Science and Electrical and Computer Engineering in the King Abdullah University of Science and Technology (KAUST) Visual Computing Center, for which he also served as director from 2012 to 2021. Prof. Heidrich joined KAUST in 2014, after 13 years as a faculty member at the University of British Columbia. He received his Ph.D. from the University of Erlangen in 1999, and then worked as a Research Associate in the Computer Graphics Group of the Max-Planck Institute for Computer Science in Saarbrucken, Germany, before joining UBC in 2000. Prof. Heidrich’s research interests lie at the intersection of imaging, optics, computer vision, computer graphics, and inverse problems. His more recent interest is in computational imaging, focusing on hardware-software co-design of the next generation of imaging systems, with applications such as High-Dynamic Range imaging, compact computational cameras, hyperspectral cameras, to name just a few. Prof. Heidrich’s work on High Dynamic Range Displays served as the basis for the technology behind Brightside Technologies, which was acquired by Dolby in 2007. Prof. Heidrich is a Fellow of the IEEE and Eurographics, and the recipient of a Humboldt Research Award.
\end{IEEEbiography}

\begin{IEEEbiography}[{\includegraphics[width=1in,height=1.25in,clip,keepaspectratio]{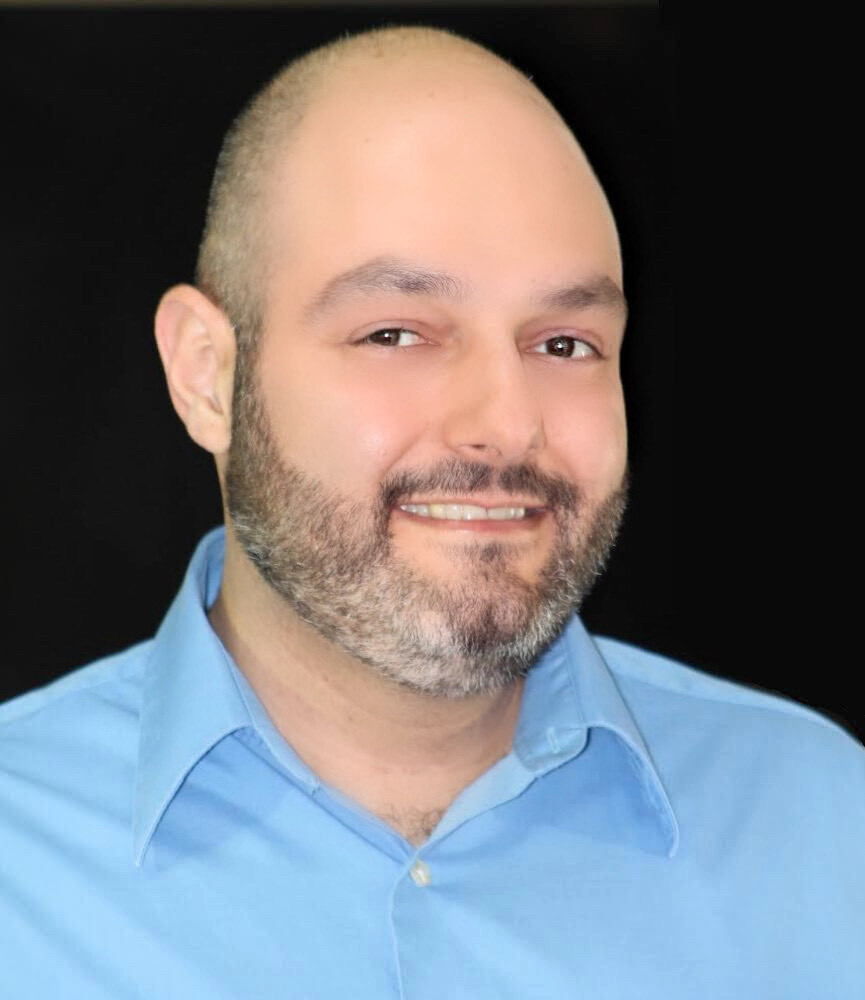}}]{Bernard Ghanem} is currently a Professor in the CEMSE division, a theme leader at the Visual Computing Center (VCC), and the Deputy Director of the AI Initiative at King Abdullah University of Science and Technology (KAUST). His research interests lie in computer vision and machine learning with emphasis on topics in video understanding, 3D recognition, and theoretical foundations of deep learning. He received his Bachelor's degree from the American University of Beirut (AUB) in 2005 and his MS/PhD from the University of Illinois at Urbana-Champaign (UIUC) in 2010. His work has received several awards and honors, including six Best Paper Awards for workshops in CVPR, ICCV, and ECCV, a Google Faculty Research Award in 2015 (1st in MENA for Machine Perception), and a Abdul Hameed Shoman Arab Researchers Award for Big Data and Machine Learning in 2020. He has co-authored more than 150 peer reviewed conference and journal papers in his field as well as three issued patents. He serves as an Associate Editor for IEEE Transactions on Pattern Analysis and Machine Intelligence (TPAMI) and has served several times as Area Chair (AC) for CVPR, ICCV, ECCV, ICLR, AAAI, and NeurIPS.
\end{IEEEbiography}

\begin{IEEEbiography}[{\includegraphics[width=1in,height=1.25in,clip,keepaspectratio]{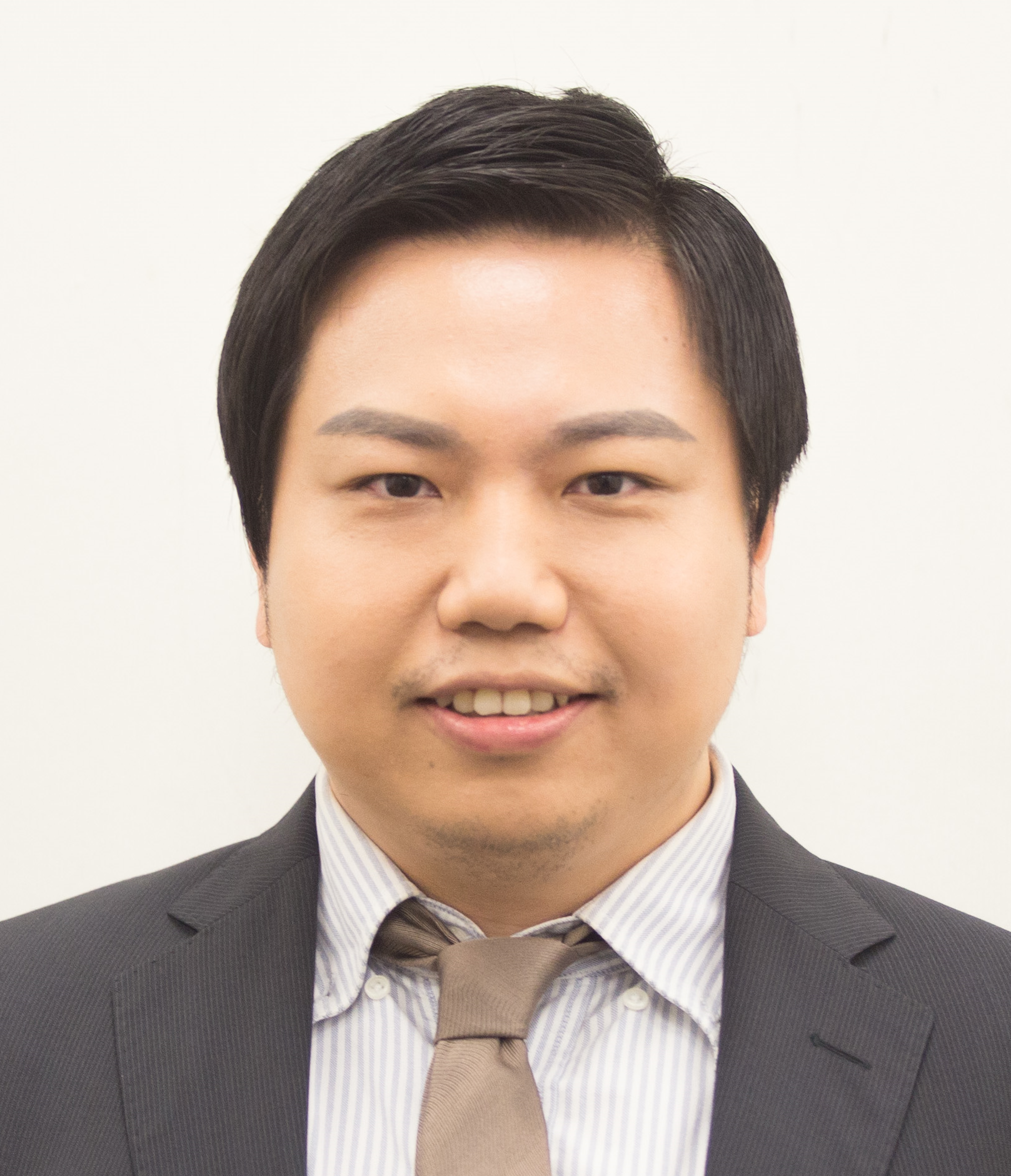}}]{Jimmy S. Ren} is currently a senior research director at SenseTime where he leads a team to build high impact computational photography products. He also holds an adjunct faculty position in Qing Yuan Research Institute, Shanghai Jiao Tong University. He received his Ph.D. degree from City University of Hong Kong in 2013. His research interests are computational photography, image processing and computer vision.
\end{IEEEbiography}




\fi

\end{document}